\documentclass[a4paper,twocolumn,showpacs,superscriptaddress,floatfix]{revtex4-1}
\usepackage{graphicx}
\usepackage{epsf}
\usepackage{epsfig}
\usepackage{amssymb,amsmath}
\usepackage[usenames]{color}
\usepackage{amssymb}
\usepackage{times}
\usepackage[czech]{babel}

\newcommand{\be}{\begin{equation}}
\newcommand{\ee}{\end{equation}}
\newcommand{\bea}{\begin{eqnarray}}
\newcommand{\eea}{\end{eqnarray}}
\newcommand{\nn}{\nonumber}

\newcommand{\hajicek}{H\'aji\v{c}ek}

\font\tenscr=rsfs10 scaled1100
\font\sevenscr=rsfs7 
\font\fivescr=rsfs5 
\skewchar\tenscr='177
\skewchar\sevenscr='177
\skewchar\fivescr='177
\newfam\scrfam
\textfont\scrfam=\tenscr
\scriptfont\scrfam=\sevenscr
\scriptscriptfont\scrfam=\fivescr

\begin{document}

\title{
Black hole horizons and quantum charged particles
}

\author{Jos\'e Luis Jaramillo}
\affiliation{
Universit\'e de Bretagne Occidentale, Brest, France}

\begin{abstract}
We point out a structural similarity between the characterization
of black hole apparent horizons as stable marginally outer trapped surfaces 
(MOTS)
and the quantum description of a non-relativistic charged particle moving in
given magnetic and electric fields on a closed surface.
Specifically, the spectral problem of the MOTS-stability operator 
corresponds to a stationary quantum particle with a formal fine-structure 
constant  $\alpha$ of negative sign.
We discuss how such analogy enriches both problems, 
illustrating this with the insights into the  
MOTS-spectral problem gained from the analysis
of the spectrum of the  quantum charged particle Hamiltonian.

\end{abstract}


\maketitle

\section{Introduction: a formal analogy}
Analogies between physical systems, either of mathematical or 
physical nature, often play a fundamental
catalyst role in conceptual and/or technical developments of the respective
theories~\cite{JonaLasinio:2010rt}. 
We discuss here a mathematical analogy between the descriptions of
black hole horizons and quantum charged particles, 
that opens a domain of cross-fertilization between quantum mechanics and
gravitation theory. More specifically, apparent horizons
--namely {\em marginally outer trapped
surfaces} (MOTS)-- possess a stability notion that guarantees 
their physical consistency as models of black hole horizons. 
Such MOTS-stability notion
\cite{AndMarSim05} 
admits a spectral characterization in terms of the so-called
{\em principal} eigenvalue of the operator
\bea
\label{e:MOTS_stability_operator}
L_{\cal S} = -\Delta  + 2 \Omega^a  D_a
- \left( |\Omega|^2
- D_a  \Omega^a  -\frac{1}{2}R_{\cal S} + G_{ab}k^a\ell^b \right) \ \ ,
\eea
defined on the apparent horizon ${\cal S}$. The terms modifying the
Laplacian $\Delta$ on ${\cal S}$ are determined by the
intrinsic and extrinsic geometry of the apparent horizon and the 
gravitational equations via the Einstein tensor $G_{ab}$ (see next
section for details).
The relevant remark in the present context is that under the 
complexification of the vector $\Omega^a$ and the identifications
\bea
\label{e:MOTS_QuantumParticle_Analogy}
\Omega_a \leftrightarrow  \frac{ie}{\hbar c}A_a \ \ , \ \ R_{\cal S} 
\leftrightarrow \frac{4me}{\hbar^2}\phi
\ \ , \ \ G_{ab}k^a\ell^b \leftrightarrow -\frac{2m}{\hbar^2}V \ \ ,
\eea
the MOTS-stability operator becomes 
$\frac{\hbar^2}{2m}L_{\cal S} \leftrightarrow \hat{H}$, 
where
\bea
\label{e:quantum_Ham}
\hat{H} &=& -\frac{\hbar^2}{2m}\Delta  + \frac{i\hbar e}{mc} A^a D_a
+ \frac{i\hbar e}{2mc} D_a A^a +  \frac{e^2}{2mc^2} A_a A^a \nn \\
&&+ e \phi + V = \frac{1}{2m}\left(-i\hbar D - \frac{e}{c} A\right)^2 
+ e\phi + V \ \ ,
\eea
is the Hamiltonian of a non-relativistic particle with mass $m$ 
and charge $e$ moving on ${\cal S}$ under magnetic and electric 
fields with vector and scalar potentials
given by $A^a$ and $\phi$, and an external potential $V$.
This formal mathematical analogy relies on a 
simple but crucial remark: 
the derivative and $\Omega^a$ terms 
in $L_{\cal S}$ can be collected in a perfect square as follows
\bea
\label{e:MOTS_stability_operator_v2}
L_{\cal S} \psi = 
\left[-\left(D - \Omega\right)^2 +\frac{1}{2}R_{\cal S} 
- G_{ab}k^a\ell^b \right]\psi \ \ .
\eea
Beyond its aesthetic appeal, and in spite of the key
difference in the self-adjoint nature of the operators, 
this analogy has the potential to open bridges 
between the well-studied quantum particle problem and 
the rich but largely uncharted MOTS subject, with applications
ranging from the MOTS-spectral problem to the spinorial
formulation of MOTS stability.
We focus here on the study of the full $L_{\cal S}$ spectrum, 
a challenging problem formulated in \cite{Jaramillo:2013rda}
in the setting of a black hole/fluid 
analogy, but with a definite geometric interest on its own.

\section{Geometry and stability of MOTS}
\subsection{Some MOTS geometry}
Let us consider a codimension-$2$ surface ${\cal S}$,
spacelike, closed (compact and without boundary)
and orientable, embedded in a $n$-dimensional spacetime $(M, g_{ab})$.
The spacetime
Levi-Civita connection is denoted by $\nabla_a$ with
Einstein curvature tensor $G_{ab}$. Let us denote 
the induced metric on ${\cal S}$ by $q_{ab}$, with Levi-Civita
connection $D_a$, Ricci scalar $R_{\cal S}$, volume form
$\epsilon_{ab}$ and measure $\eta_{\cal S}$.
Let us consider future-oriented null vectors $\ell^a$ and $k^a$ spanning 
the normal bundle $T^\perp{\cal S}$ and normalized as 
$\ell^a k_a = -1$. 
This normalization leaves a null-vector-rescaling freedom by a positive
function $f>0$ respecting time orientation, 
corresponding to a boost transformation 
\bea
\label{e:gauge_freedom}
\ell'^a =f \ell^a \ \ , \ \ k'^a = f^{-1} k^a \ .
\eea
We define the expansion $\theta^{(\ell)}$ and \hajicek{} or rotation
form $\Omega_a$
\bea
\label{e:theta_Omega}
\theta^{(\ell)}\equiv q^{ab}\nabla_a\ell_b \ \ \ , \ \ \ 
\Omega_a\equiv -k^c {q^d}_a \nabla_d \ell_c \ ,
\eea
associated with $\ell^a$. 
Considering 
$v^a = \alpha \ell^a + \beta k^b$, 
we can write ${q^c}_a\nabla_c v_b =  D_a^\perp v_b + \Theta_{ab}^{(v)}$ , with
$\Theta_{ab}^{(v)}\equiv{q^c}_a {q^d}_b \nabla_c v_d$ and
\bea
\label{e:normalconnection1}
D_a^\perp v_b = 
(D_a \alpha + \Omega_a \alpha )\ell_b + (D_a \beta - \Omega_a \beta )k_b \ .
\eea
The \hajicek{} form therefore provides a connection
on the normal bundle $T^\perp {\cal S}$ for the tangent derivative of normal vectors.
From a physical perspective it represents a sort of angular
momentum density. Given an axial Killing vector $\phi^a$ on ${\cal S}$,
\bea
\label{e:angular_momentum}
J[\phi]=\frac{1}{8\pi}\int_{\cal S} \Omega_a \phi^a \eta_{\cal S}
\eea
is the (Komar) angular momentum associated with ${\cal S}$. 
Regarding the
expansion, the surface  ${\cal S}$ is a marginally outer trapped surface (MOTS)
if it satisfies the condition: $\theta^{(\ell)}=0$.
We refer then  
to $\ell^a$ as {\em outgoing} and to 
$k^a$ as {\em ingoing} null vectors.

\subsection{MOTS stability and MOTS-stability operator}
A MOTS ${\cal S}$ is said to be {\em stable} (more properly, stably outermost 
\cite{AndMarSim05,AndMarSim07}) 
in the ingoing $k^a$ direction
if it can be infinitesimally deformed along $k^a$ into a  properly 
(outer) trapped surface ${\cal S}'$, i.e. with $\theta^{(\ell)}|_{{\cal S}'} < 0$. 
Using the deformation operator $\delta_v$ along a normal vector $v^a$ 
(discussed in \cite{AndMarSim05}), this amounts 
to the existence of a function $\psi>0$ such that $\delta_{\psi k}\theta^{(\ell)}<0$.
Such a condition admits a spectral characterization
in terms of the elliptic operator $L_{\cal S}$ defined as 
$L_{\cal S} \psi \equiv \delta_{\psi (-k)} \theta^{(\ell)}$, 
with explicit expression (\ref{e:MOTS_stability_operator}). 
The operator $L_{\cal S}$, namely the MOTS-stability operator, 
is generically non-selfadjoint [in $L^2({\cal S},\eta_{\cal S})]$
due to the $2\Omega^a D_a$ term. Therefore, in the eigenvalue problem
\bea
\label{e:MOTS_spectral_problem}
L_{\cal S} \psi = \lambda \psi \ ,
\eea
the $\lambda$'s are generically complex.
Their real part is bounded below, leading 
to the definition of the {\em principal eigenvalue} $\lambda_o$ 
of $L_{\cal S}$ as that with smallest real part. Lemmas 1 and 2 in 
\cite{AndMarSim05} state that: i) $\lambda_o$ is real, 
and ii) ${\cal S}$
is stably outermost iff $\lambda_o \geq 0$.

\subsection{MOTS-gauge symmetry}
The MOTS geometry described above
does not depend on the choice of null normals
(subject to $\ell^ak_a=-1$).
In this sense, the null vector rescaling freedom
(\ref{e:gauge_freedom}) is a gauge transformation of the MOTS geometry. 
We consider now the transformation of the main objects on ${\cal S}$ under
the rescaling (\ref{e:gauge_freedom}).

{\bf Lemma 1} (MOTS-gauge transformations).   
{\em Under the null normal rescaling $\ell'^a= f \ell^a$, $k'^a= f^{-1} k^a$, 
with $f>0$:
\begin{itemize}
\item[i)] The expansion and \hajicek{} form transform as
\bea
\label{e:theta_Omega_transformations}
\theta^{(\ell')}=f \theta^{(\ell)} \ \ \ , \ \ \ \Omega'_a = \Omega_a + D_a(\mathrm{ln}f) \ .
\eea
\item[ii)] The MOTS-stability operator transforms covariantly
\bea
\label{e:transformation_L_S}
(L_{\cal S})' \psi  = f L_{\cal S} (f^{-1}\psi) \ ,
\eea
where $(L_{\cal S})'\psi \equiv  \delta_{\psi (-k')} \theta^{(\ell')}$.
\item[iii)] The MOTS-eigenvalue problem is invariant under
the additional  eigenfunction transformation, $\psi'= f\psi$
\bea
\label{e:transformation_spectral_problem}
L_{\cal S}\psi = \lambda \psi \ \ \to \ \ (L_{\cal S})' \psi'  = \lambda \psi' \ .
\eea
\end{itemize}
}

{\em Proof:} Point {\em i)} follows directly by plugging (\ref{e:gauge_freedom})
into (\ref{e:theta_Omega}). Regarding point {\em ii)},
although it can be obtained by straightforward substitution
of (\ref{e:theta_Omega_transformations}) into (\ref{e:MOTS_stability_operator}),
it is simpler to use the definition of $L_{\cal S}$. Considering its
action on a function $\psi$
\bea
\label{e:L_k_transformation_2}
(L_{\cal S})'\psi &=&  \delta_{\psi (-k')} \theta^{(\ell')} 
= \delta_{\psi (-k')} (f\theta^{(\ell)})  \\
&=&\delta_{\psi (-k')} (f)\theta^{(\ell)}+
f \delta_{\psi (-k')} \theta^{(\ell)} = f \delta_{\psi (-k')} \theta^{(\ell)} \nn \\
&=&
f \delta_{\psi (-f^{-1}k)} (\theta^{(\ell)}) = f \delta_{(f^{-1}\psi) (-k)} \theta^{(\ell)} =
f L_{\cal S} (f^{-1}\psi)  \nn
\eea
where in the first line we have used the $\theta^{(\ell)}$ transformation 
in (\ref{e:theta_Omega_transformations}), the second line uses the Leibnitz rule holding
for $\delta_v$ and the MOTS condition, and in the third line
we used again the definition of $L_{\cal S}$. Finally, point {\em iii)}
follows directly
\bea
(L_{\cal S})'\psi' = f L_{\cal S} (f^{-1}\psi') = f L_{\cal S} (\psi)
=  f \lambda \psi = \lambda \psi' \ .
\eea

\medskip
Point {\em i)} just states the invariance of the MOTS notion under
(\ref{e:gauge_freedom}) and the transformation of 
$\Omega_a$ as a connection
under the (multiplicative) abelian gauge group $\mathbb{R}^+$ of 
positive null rescalings. The latter is
consistent with the nature of $\Omega_a$ in (\ref{e:normalconnection1}) 
as a connection in the normal bundle. Point {\em ii)}, stating the good 
(covariant) transformation
properties of $L_{\cal S}$ under $\mathbb{R}^+$, is the analogue in the present 
setting of Proposition 4 in \cite{Mars:2012sb} concerning 
the free choice of section of stationary black hole horizons. 
Point {\em iii)} guarantees that the MOTS-eigenvalue problem is well-defined
and provides the gauge transformation rule for the associated eigenfunctions.
Of course, all these points evoke familiar features of the quantum charged particle.

\section{MOTS and quantum charged particles}
To take a step further from the formal correspondance 
(\ref{e:MOTS_QuantumParticle_Analogy}) into a more precise
statement, let us review the stationary
quantum charged particle (QCP) problem.
The Schr\"odinger equation for a non-relativistic 
(spin-$0$) charged particle moving in 
electromagnetic 
fields with magnetic vector potential $A^a$ and electric potential $\phi$,
namely $i\hbar \partial_t \Psi = \hat{H}\Psi$ [with $\hat{H}$
in (\ref{e:quantum_Ham})], follows from that of 
a non-charged particle in an external mechanical potential $V$ 
via a minimal-coupling prescription
\bea
\label{e:minimal_coupling}
i\hbar \partial_t \to i\hbar \partial_t - e \phi \ \ , \ \ 
-i\hbar D_a \to -i\hbar D_a -\frac{e}{c}A_a \ .
\eea
The stationary equation for the energy eigenvalues $E$ is then
\bea
\label{e:stationary_Schrodinger}
\left[\frac{1}{2m}(-i\hbar D_a -\frac{e}{c}A_a)^2 
+ e\phi + V\right]\psi = E \psi \ ,
\eea
where $\Psi = e^{-iEt/\hbar} \psi$ (with $\partial_t \psi = 0$).
This equation should not depend on the gauge choice
of the electromagnetic potentials. The gauge
transformation of $A_a$ by a total gradient~\footnote{The electric
potential $\phi$ stays invariant $\phi \to \phi - 
\frac{1}{c}\partial_t \sigma$ under gauge transformations compatible
with stationarity,  $\partial_t \sigma=0$.}
\bea
\label{e:electromagnetic_gauge_transformations}
A_a \to A_a - D_a \sigma \ ,
\eea 
leaves Eq. (\ref{e:stationary_Schrodinger}) invariant
if we simultaneously transform $\psi$ as
\bea
\label{e:wave_function_transformation}
\psi \to e^{ie\sigma/(c\hbar)} \psi \ \ ,
\eea
i.e. by a (local) phase. 
Transformations (\ref{e:electromagnetic_gauge_transformations})
and (\ref{e:wave_function_transformation}) define 
the electromagnetic abelian $U(1)$-gauge symmetry.
From these remarks we can state the following
similarities between the eigenvalue problems
(\ref{e:MOTS_spectral_problem}) and 
(\ref{e:stationary_Schrodinger}), placing the
MOTS-QCP analogy in 
(\ref{e:MOTS_QuantumParticle_Analogy}) on a 
sounder structural basis~\footnote{
A further point could be the understanding of MOTS-stability, 
$\lambda_o\geq 0$, as a 
MOTS-counterpart of a positivity condition on the quantum ground
state $E_o$, refining 
quantum stability. This is however delicate,
since the operator correspondance (\ref{e:MOTS_QuantumParticle_Analogy}) 
does not necessarily preserve eigenvalue signs
(see section \ref{s:MOTSspectralproblem}).
}:

{\em i) Abelian gauge symmetry}. The QCP 
eigenvalue problem (\ref{e:stationary_Schrodinger}) and the 
MOTS-spectral problem (\ref{e:MOTS_spectral_problem}) are respectively
invariant under transformations (Eqs. (\ref{e:electromagnetic_gauge_transformations})-(\ref{e:wave_function_transformation}) and Lemma 1)
\be
\label{e:compared_gauge_transformations}
\begin{array}{rcclcl}
\hbox{QCP:} & A_a &\to& A_a - D_a \sigma & , &  \psi \to e^{ie\sigma/(c\hbar)} \psi  \\
\hbox{MOTS:} &  \Omega_a &\to& \Omega_a - D_a \sigma & , &  \psi \to e^{-\sigma} \psi \ .\\ 
\end{array}
\ee
They both define
abelian symmetries of gauge nature:
in the QCP case it is
the electromagnetic $U(1)$-gauge transformation 
($g''= g\cdot g'$, with $g=e^{ie\sigma/(c\hbar)}$) relying on the 
phase invariance of the wave function,
whereas for MOTSs it defines a non-compact 
$\mathbb{R}^+$-gauge counterpart ($g''= g\cdot g'$, with $g\!=\!f=\!e^{-\sigma}$) 
reflecting the in-built
null rescaling (boost) freedom of the MOTS geometric description.
In brief,
the MOTS-spectral problem presents symmetry transformation
properties in full analogy with those of the QCP Schr\"odinger equation. 

{\em  ii) Gauge field potential}.
In this symmetry setting, the 1-form $\Omega_a$ 
emerges as the natural gauge field of the $\mathbb{R}^+$-gauge
group. This endorses, at a structural level, its purely 
formal correspondence in 
(\ref{e:MOTS_QuantumParticle_Analogy}) with the $A_a$ magnetic 
$U(1)$-gauge field. We note that the normal connection 
in (\ref{e:normalconnection1})
admits an interpretation as a gauge connection:
$\ell^b D_a^\perp (\psi k_b) = -(D_a -\Omega_a)\psi$.

{\em  iii) Minimal Coupling.}
It is at a ``dynamical'' level
where the analogy proves remarkable: 
the $\Omega_a$ field enters in $L_{\cal S}$ via a standard
gauge ``minimal coupling'' mechanism, namely a shift in the 
Levi-Civita connection with the gauge connection
\bea
\label{e:minimal_coupling_Omega}
D_a \to D_a - \Omega_a \ .
\eea
This becomes
apparent in the perfect-square version (\ref{e:MOTS_stability_operator_v2})
of $L_{\cal S}$. Therefore, in full analogy with the  minimal coupling
mechanism for incorporating 
the magnetic field in the QCP problem via the shift
(\ref{e:minimal_coupling}) in the non-charged equations, 
rotation in a MOTS is switched-on via the minimal coupling 
(\ref{e:minimal_coupling_Omega}). 

\subsection{MOTSs and a negative ``fine structure constant'' $\alpha$}
Setting 
$\hbar\!\!=\!\!m\!\!=\!\!c\!\!=\!\!1$ 
and introducing a formal complex 
``fine-structure constant'' $\alpha\equiv e^2$,
we define the operator family 
\bea
\label{e:L[alpha]}
\!\!\!\!\!\! &&L[\sqrt{\alpha}] = -\frac{1}{2}(D -i\sqrt{\alpha}\Omega)^2 
- \frac{\alpha}{4}R_{\cal S} - \frac{1}{2}G_{ab}k^a\ell^b \\ 
\!\!\!\!\!\!&=&  \frac{-1}{2}\Delta + i\sqrt{\alpha}(\Omega\cdot D
+ \frac{1}{2}D\cdot \Omega) + \frac{\alpha}{2}|\Omega|^2 
- \frac{\alpha}{4}R_{\cal S} - \frac{1}{2}G(k,\ell). \nn
\eea
The QCP Hamiltonian corresponds to the (normalized) standard real positive 
$\alpha=1$, whereas (half) the MOTS-stability operator corresponds
to a negative $\alpha=-1$. Specifically, QCP and MOTS operators
are recovered with branch choices: $\hat{H}=L[\sqrt{\alpha}=1]$
and $L_{\cal S}/2=L[\sqrt{\alpha}=-i]$.
In this sense, stable MOTSs can be seen as QCPs
with negative ``fine-structure constant'' $\alpha$.  
This suggests to import QCP terms to MOTSs.

{\em Terminology for $L_{\cal S}$ terms.} 
Regarding terms containing the rotation field $\Omega_a$,
we refer to $|\Omega|^2$ 
as the {\em diamagnetic} term, whereas $\Omega\cdot D$ is
the {\em paramagnetic} term \cite{GalPasII91}. The divergence 
$D\cdot\Omega$ 
is a {\em gauge-fixing} term and can be chosen 
by an appropriate transformation 
(\ref{e:compared_gauge_transformations}).
For completeness sake,
$\Delta$ is the
{\em kinematical} term and $G(k,\ell)$ is the
{\em external mechanical} potential.
Finally, $R_{\cal S}/4$ can be referred to as the
{\em electric} potential term. To justify its explicit distinction
from the {\em external mechanical} potential, we consider 
in the $2$-dimensional case the complex scalar 
${\mathcal K}$  on ${\cal S}$ introduced by Penrose and Rindler 
\cite{PenRin86} as
\bea
{\mathcal K} = \frac{1}{4}R_{\cal S} + i \frac{1}{4}\epsilon^{ab}F^\Omega_{ab} \ ,
\eea
where $F^\Omega_{ab}=D_a\Omega_b -D_b\Omega_a$, namely the curvature of 
$\Omega_a$. The real and imaginary parts of ${\mathcal K}$
correspond, respectively, to
{\em electric} and {\em magnetic} terms.
This gravity/electromagnetic analogy in ${\mathcal K}$ has been 
used to discuss isolated/dynamical horizon
source multipoles \cite{AshEngPaw04} and to introduce the notions of
``vortexes'' and ``tendexes'' in the analysis of dynamical black 
holes \cite{OweBriChe11}.
The present discussion promotes such analogy to a sounder structural level
by identifying the symmetry and minimal coupling similarities in the
relevant operators.

The ultimate motivation behind this analogy
is to explore the transfer of concepts and tools
between both problems.
This can prove fruitful for the MOTS-spectral problem
by profiting of the extensive knowledge accumulated about
QCP bound states. In parallel, the development
of spinor treatments of MOTS-stability can largely benefit from the
presented analogy. In particular, using the Lichnerowicz-Weitzenb\"ock
formula to mimic Pauli's approach to spin 
can provide insights in the structure of the MOTS 
second-order operator $L_{\cal S}$,
whereas Dirac's approach to spin can open an avenue to a 
first-order formulation of MOTS-stability (of potential
interest in boundary value problems of bulk spinor equations). 
The GHP formalism \cite{gerochheld73:_ghp} can offer additional insights
in this setting.
We postpone the development of spinor approaches 
to future studies and focus in the following on the MOTS-spectral problem.

\section{MOTS-spectral problem}
\label{s:MOTSspectralproblem}

\subsection{An explicit example: ``MOTS-Landau'' levels}
\label{s:explicitexample}
We consider now the behaviour of the eigenvalues and
eigenfunctions in problems 
(\ref{e:stationary_Schrodinger}) and (\ref{e:MOTS_spectral_problem}), 
under the complex rotation (from $\sqrt{\alpha}=1$ to $\sqrt{\alpha}=-i$, 
in $L[\sqrt{\alpha}]$) 
that realizes the analogy (\ref{e:MOTS_QuantumParticle_Analogy}).
In this context, and following the spirit of Landau levels of 
a QCP moving in a constant magnetic field (that provides an 
explicit example illustrating basic features of such quantum systems),
we start by discussing a simple eigenvalue problem for $L_{\cal S}$ that
can be explicitly solved both in the MOTS case and, independently, in the
analogous QCP case.

Let us take a 
2-sphere ${\cal S}=S^2$ with ``round'' metric 
$q_{ab}=r^2(d\theta^2 + \sin^2 \theta d\varphi^2)$.
Decomposing
the \hajicek{} form as
$\Omega_a = {\epsilon_a}^b D_b \omega + D_a \zeta$ with the
simplest non-trivial choice $\omega= a \sin \theta, \zeta =0$ 
($a\in \mathbb{R}$), 
we have:
$\Omega = a \sin^2 \theta d\varphi$. 
In vacuum (i.e. $G_{ab}=8\pi T_{ab}=0$), the solution to the 
MOTS-eigenvalue problem for the resulting $L_{\cal S}$ 
is given explicitly by
\bea
\label{e:Landau_MOTS}
\!\!\!\!\!\!\!\!\!\!\!\!\! \lambda = \frac{1}{r^2}\left[(\lambda_{\ell m}(a) 
+ 1 - a^2) + i 2 a m\right]  , 
\psi = S_{\ell m}(a,\cos\theta) e^{im\varphi} 
\eea
where $S_{\ell m}(a,\cos\theta)$ are the {\em prolate} spheroidal functions
with eigenvalues $\lambda_{\ell m}(a)$ (21.6.2 in \cite{AbramowitzStegun64}).
Standard spherical harmonics are recovered
in the limit $a\to 0$:
$\lambda_{\ell m}(a)\to \ell(\ell+1)$ and  $S_{lm}(a,\cos\theta) \to 
P_{\ell m}(\cos \theta)$). We can now consider the ``QCP 
counterpart'' by performing the complex rotation $a\to i a$ 
in the operator $L_{\cal S}$. The resolution of the new
eigenvalue problem leads to
eigenvalues $\bar{\lambda}$ and 
eigenfunctions $\bar{\psi}$ 
\bea
\label{e:Landau_QCP}
\!\!\!\!\!\!\!\!\!\!\!\!\!\bar{\lambda} = 
\frac{1}{r^2}\left(\bar{\lambda}_{\ell m}(a) + 1 + a^2 
- 2 a m 
\right)   \  ,  \
\bar{\psi} = \bar{S}_{\ell m}(a,\cos\theta) e^{im\varphi} 
\eea
where $\bar{S}_{\ell m}(a,\cos\theta)=S_{\ell m}(ia,\cos\theta)$
are now the {\em oblate} spheroidal functions with eigenvalues 
$\bar{\lambda}_{\ell m}(a)=\lambda_{\ell m}(ia)$ (cf. 21.6.4 and 21.7.5
in \cite{AbramowitzStegun64}; note $\lambda_{\ell m}(ia)\in\mathbb{R}$). 
Therefore, at least in this simple example it is verified 
that  eigenvalues $\lambda $ 
(eigenfunctions $\psi$) of the 
operator $L_{\cal S}=2L[-ia]$~\footnote{$L[-ia]$ assumes implicitly 
a H\'aji\v {c}ek{} form  $\Omega = \sin^2 \theta d\varphi$.} 
 can be actually recovered
by solving for the ``rotated'' $a\to ia$ self-adjoint operator 
$L[a]$, and then inverting the rotation
$a\to \frac{1}{i}a=-ia$ in the resulting eigenvalues $\bar{\lambda}$
(eigenfunctions $\bar{\psi}$).

\subsection{Analyticity in the ``fine structure constant''} 
The fact that the MOTS-stability operator can be obtained 
from the QCP Hamiltonian as an analytic continuation of $L[\sqrt{\alpha}]$
($\sqrt{\alpha}=1\to\sqrt{\alpha}=-i$), together with the 
discussion of the previous explicit example, raise the following 
question: {\em can we recover the MOTS-spectrum ($\alpha=-1$) as an 
analytic extension of the QCP spectrum ($\alpha=1$) self-adjoint problem?} 

This question dwells naturally in the perturbation
theory of linear operators (where $L[\sqrt{\alpha}]$
defines a self-adjoint holomorphic family of type (A) \cite{Kato80}),
but giving a fully general answer defines a difficult problem.
A given eigenvalue $\lambda(\sqrt{\alpha})$ can be analytically 
continued along its path in the complex plane, as long as 
its  evolution does not encounter (for the same $\sqrt{\alpha}$)
another eigenvalue. 
But checking this is a hard task even in the explicit example
above. On the other hand, our particular setting is free of two 
potential threats for the analyticity discussion,
namely boundaries and function pathologies:
{\em i)} ${\cal S}$ has no boundaries 
(is closed), and {\em ii)} the functions in $L[\sqrt{\alpha}]$, being induced
from the ambient geometry, can be taken as regular as needed. 
As a third point {\em iii)}, potential topological issues
associated to the underlying $U(1)$ or $ \mathbb{R}^+$-fibre bundle are absent
since such bundle is trivial (we are excluding here the possibility
of a non-trivial NUT charge). Supported by these points and
in the assumption that the example in \ref{s:explicitexample}
contains all the relevant qualitative elements, we propose the following:

\medskip
\noindent {\bf Analyticity Conjecture.}  
{\em Given an orientable closed surface ${\cal S}$
and the one-parameter family of operators $L[\sqrt{\alpha}]$ defined 
in (\ref{e:L[alpha]}),
the MOTS-spectrum 
($\sqrt{\alpha} = -i$) can be recovered as an ``analytic continuation''
of the QCP spectrum ($\sqrt{\alpha} = 1$).
} 
\medskip

We present this conjecture as an open problem.
In case the conjecture proves to be valid~\footnote{Even without 
a continuation argument, the prescription 
$\protect \sqrt {\alpha }\to -i\protect \sqrt {\alpha }$
to recover MOTS spectra from QCPs could still hold.}, the
MOTS-stability spectrum problem would be ``essentially'' reduced to that 
of the self-adjoint problem of the stationary non-relativistic QCP.


\subsection{Ground state of the charged particle}
As a first application, we consider a transfer in the ``inverse'' sense,
by using a MOTS result to calculate the
ground state energy $E_o$ of QCPs. In \cite{AndMarSim07} a variational
Rayleigh-Ritz-like expression for $\lambda_o$ is presented.
This remarkable result does not follow from the 
Rayleigh-Ritz characterization, since $L_{\cal S}$ is generically 
not selfadjoint. The expression
for $\lambda_o$ is rather obtained by starting from a 
min-max characterization by Donsker and Varadhan \cite{DonVar75},
valid for real not necessarily selfadjoint operators. If the conjecture
above proves true, the ``rotation'' $\sqrt{\alpha}\to -i\sqrt{\alpha}$
in the $\lambda_o$ of \cite{AndMarSim07} results in a  QCP $E_o$ 
\bea
\label{e:E_0_gaugeinvariant}
E_o = \underset{\psi>0}{\mathrm{inf}} \int_{\cal S} \left(
|D\psi|^2 + \left(e\phi + V +e^2|D\omega_\psi + z|^2\right) \psi^2 
\right) \eta_{\cal S}
\eea
where $A_a = z_a + D_a \zeta$ (with $D_a z^a=0$), 
$\int_{\cal S} \psi^2 \eta_{\cal S} = 1$ and
$\omega_\psi$ satisfies, for a given $\psi>0$, the constraint equation
\bea
\label{e:E_0_constraint}
-\Delta \omega_\psi -\frac{2}{\psi} D_a \psi D^a\omega_\psi = \frac{2}{\psi} z^a D_a\psi \ \ .
\eea 
This expression for $E_o$, ``blindly'' transported from the 
MOTS result, has two virtues as compared with the
straightforward evaluation of the Rayleigh-Ritz 
$E_o= \underset{||\psi||=1}{\mathrm{inf}} \int_{\cal S} \psi^* \hat{H} \psi\eta_{\cal S}$: i) it is explicitly gauge-invariant, since the term
$D_aA^a(=\Delta \zeta)$ is absent; and ii) the paramagnetic term is 
recast as a diamagnetic one, something of potential interest in 
numerical strategies. Then, a crucial point is that the
``blind'' expression (\ref{e:E_0_gaugeinvariant}) for QCPs can actually 
be proved: starting from the (now valid) Rayleigh-Ritz expression 
and adapting the steps in \cite{AndMarSim07}
to the MOTS/QCPs analogy, expression (\ref{e:E_0_gaugeinvariant})
follows. This is remarkable since, although it is known \cite{DonVar75} that
Rayleigh-Ritz and Donsker-Varadhan expressions coincide when they both apply
(namely, selfadjoint real operators), they cannot be generically
reduced to one another (in particular in our
setting $A_a\neq 0$). 
Therefore, the fact that  (\ref{e:E_0_gaugeinvariant}) still holds 
in the fully general case
offers a first non-trivial test of the conjecture.

\subsection{Semi-classical approach to the MOTS spectrum}
As a second example, the effective reduction to a selfadjoint 
problem would open a particular
avenue to the use of approximate tools in the MOTS
spectral problem, by transferring the semi-classical tools 
for QCPs \footnote{But note that ``non-selfadjoint'' 
semi-classical tools also exist.}.
More specifically, starting from the MOTS-stability operator,
we can in a first step consider the analogous QCP  Hamiltonian
$\hat{H}[\sqrt{\alpha}\in \mathbb{R}]$, and in a second
step its corresponding classical Hamiltonian function
\bea
\label{e:classical_Ham}
H_{\mathrm{cl}}[\sqrt{\alpha}](x,p) = (p - \sqrt{\alpha}\Omega)^2 
+ \frac{1}{2}R_{\cal S} - G(k,\ell) \ ,
\eea
defined on the cotangent bundle $T^*{\cal S}$
by reverting the ``quantization rule'', $p_i \to -i D_i$.  
Approximate eigenvalues and eigenfunctions 
for the MOTS problem could then obtained,
assuming the validity of the conjecture,  by applying 
semi-classical tools on $H_{\mathrm{cl}}[\sqrt{\alpha}]$ and
then evaluating the explicit result on $\sqrt{\alpha}\to -i$.
Whereas WKB techniques could be appropriate in separable problems, 
generic cases (notably, $\Omega_a\neq 0$) would 
need to resort to the rich 
semi-classical tools developed in the setting of
quantum chaos studies (e.g. \cite{Berry83,Nonne10}).


\medskip

\smallskip\noindent\emph{Acknowledgments.~} 

\noindent I thank A. Afriat, C. Aldana, V. Aldaya, L. Andersson, 
M. Ansorg, A. Ashtekar, 
C. Barcel\'o, J. Bi\v{c}\'ak, 
M.E. Gabach-Cl\'ement,
Q. Hummel, A. Marquina, M. Mars, G. Mena-Marug\'an,
J.P. Nicolas, I. R\'acz, M. Reiris, L. Rezzolla,
M. S\'anchez, J.M.M. Senovilla, W. Simon, S. Nonnenmacher, J.D. Urbina
and J.A. Valiente-Kroon.
I fondly thank E. Alcal\'a, whose inspiring images made this research possible.





\end{document}